\documentclass[twocolumn,prl,amsmath,superscriptaddress,showpacs]{revtex4}
\usepackage[T1]{fontenc}
\usepackage[latin9]{inputenc}
\usepackage{amssymb}
\usepackage{graphicx}
\usepackage{subscript}

\makeatletter
\@ifundefined{textcolor}{}
{%
 \definecolor{BLACK}{gray}{0}
 \definecolor{WHITE}{gray}{1}
 \definecolor{RED}{rgb}{1,0,0}
 \definecolor{GREEN}{rgb}{0,1,0}
 \definecolor{BLUE}{rgb}{0,0,1}
 \definecolor{CYAN}{cmyk}{1,0,0,0}
 \definecolor{MAGENTA}{cmyk}{0,1,0,0}
 \definecolor{YELLOW}{cmyk}{0,0,1,0}
}

\makeatother

\usepackage[greek,english]{babel}
\begin{document}
\selectlanguage{english}

\title{Exciton Binding Energy of Monolayer WS\textsubscript{2}}

\author{Bairen Zhu\textsuperscript{+},Xi Chen\textsuperscript{+},Xiaodong
Cui}

\email{xdcui@hku.hk }

\selectlanguage{english}%

\affiliation{Department of Physics, The University of Hong Kong, Hong Kong, China}

\date{\today}
\begin{abstract}
The optical properties of monolayer transition metal dichalcogenides
(TMDC) feature prominent excitonic natures. Here we report an experimental
approach toward measuring the exciton binding energy of monolayer
WS\textsubscript{2} with linear differential transmission spectroscopy
and two-photon photoluminescence excitation spectroscopy (TP-PLE).
TP-PLE measurements show the exciton binding energy of 0.71$\pm$0.01eV
of the band-edge excitons around K valley in the Brillouin zone. 
\end{abstract}


\maketitle
Coulomb interactions are significantly enhanced in low dimensional
systems as a result of spatial confinement and reduced screening,
and consequently excitons, quasiparticles of electron-hole pairs bounded
by Coulomb force play a pronounced role in their optical aspects.
A few paradigms of the pronounced excitonic effects have been demonstrated
in quantum dots and carbon nanotubes where the exciton binding energies
are found to be a fraction of their band gaps in these quasi-zero
dimensional (0D) and one dimensional (1D) systems. Prominent exciton
effects are also widely expected in intrinsic 2D systems for instance
monolayer crystals of transition metal dichalcogenides (TMDC) owing
to the reduced dielectric screening and spatial confinement\cite{key-1,key-2}.
Monolayer TMDC is an intrinsic 2D crystal consisting of two hexagonal
planes of chalcogen atoms and an intermediate hexagonal plane of metal
atoms in a prismatic unit cell. Particularly MX\textsubscript{2}
(MoS\textsubscript{2}, MoSe\textsubscript{2}, WS\textsubscript{2}
and WSe\textsubscript{2}) experiences a transition from indirect
gap in bulk form to direct gap of visible frequency in monolayers,
where the band gap is located at K(K') valley of the Brillouin zone\cite{key-3,key-4,key-5,key-6}.
Ab initio calculations show the direct-gap exciton binding energy
in the range of 0.5-1eV which is around 1/3-1/2 of the corresponding
optical direct gap\cite{key-1,key-2,key-7,key-8}. The modulated absorption/reflection
spectroscopy shows the binding energy of direct gap excitons around
55meV in bulk crystals\cite{bulkcrystal_ref}. Such a big exciton
binding energy in bulk form guarantees the robust excitonic nature
of optical properties in ultrathin counterparts. Furthermore, photoluminescence
(PL) experiments identify electron(hole)-bounded excitons, so called
trions, with a charging energy $E_{bX^{-}}$of 18meV, 30meV and 20-40meV
in monolayer MoS\textsubscript{2}, MoSe\textsubscript{2} and WS\textsubscript{2}
respectively\cite{trion-mose2,trion_mos2,trion-WS2}. With a simple
2D exciton model, one could estimate the exciton binding energy around
10 times that of the trion, if equal effective electron's and hole's
mass are assumed\cite{trion-model}. As yet the direct measurement
of exciton binding energy in monolayer TMDC is lacking.

Here we report experimental approaches toward measuring the exciton
binding energy of monolayer WS\textsubscript{2} with linear differential
transmission spectroscopy and two-photon photoluminescence excitation
spectroscopy (TP-PLE). The TP-PLE resolves the excited states of excitons
and the interband transition continuum. The exciton binding energy
of 0.71$\pm$0.01eV of the band-edge excitons around K valley in the
Brillouin zone is extracted by the energy difference between the ground
state exciton and the onset of the interband continuum. 

Monolayer WS\textsubscript{2} was mechanically exfoliated from single
crystal WS\textsubscript{2} and identified with optical microscope
and photoluminescence spectroscopy (supplementary information). The
samples in differential transmission measurements were made by transferring
from silicon substrates to freshly cleaved mica substrates as described
in Ref\cite{transfer}. The electric gate dependent PL measurements
were carried out with a field effect transistor structure on silicon
wafers with a 300nm oxide cap layer. The TP-PLE spectroscopy was carried
out with a confocal setup with a 20X achromatic objective and a Ti:sapphire
oscillator (80MHz, 100fs). 

\begin{figure}
\includegraphics[width=8.5cm]{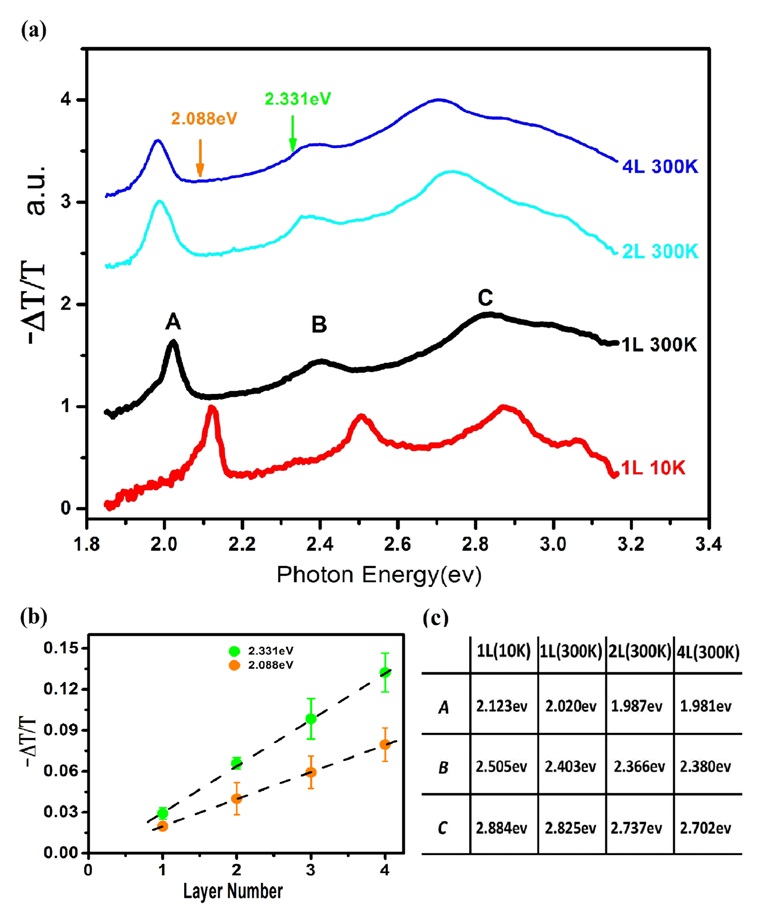}\caption{Linear absorption spectra of WS\textsubscript{2} atomically thin
films. (a) Normalized differential transmission spectra of multi-
and monolayer WS\textsubscript{2} at room temperature and 10K. (b)
Absorbance of atomically thin layer at photon energy of 2.088eV(orange)
and 2.331eV (green) respectively as indicated by arrows in figure
1(a). The absorbance shows a linearly dependence on layer number,
each unit layer with constants of 2.0\% and 3.4\% respectively. (c)
Absorption peak energy values of exciton ``A\textquotedblright{},
``B\textquotedblright{} and ``C\textquotedblright{}. }
\end{figure}

Figure 1a summarizes linear optical measurements of monolayer and
multilayer WS\textsubscript{2}. There are distinct peaks in the differential
transmission spectra, labeled as ``A\textquotedblright{}, ``B\textquotedblright{}
and ``C\textquotedblright{} respectively\cite{key-6,bulkcrystal_ref}.
Peaks ``A\textquotedblright{} around 2eV and ``B\textquotedblright{}
around 2.4eV at room temperature present the excitonic absorptions
at the direct gap located at K valley of the Brillouin zones. The
separation between ``A\textquotedblright{} and ``B\textquotedblright{}
of 0.38eV rising from the splitting of valence band minimum (VBM)
due to spin-orbit coupling (SOC) at K(K') valley is almost constant
in all the layers with various thickness, consistent with the PL spectra\cite{key-6,key-5}.
It is the direct result of the suppression of interlayer coupling
at K(K') valley owing to the giant SOC and spin-valley coupling in
tungsten TMDC with 2\textit{H} stacking order in which each unit layer
is a $\pi$ rotation of its adjacent layers\cite{key-5}. The peak
``C\textquotedblright{} around 2.8eV was recognized as the excitonic
transitions from multiple points near \foreignlanguage{greek}{$\Gamma$}\inputencoding{latin9}
point of the Brillouin zone\cite{key-2,bulkcrystal_ref}. Unlike in
many semiconductors, the linear absorption spectra of WS\textsubscript{2}
display no gap between distinct excitons and the continuum of interband
transitions. The continuous absorption originates from the strong
electron(hole)-phonon coupling in TMDCs and the efficient phonon scattering
fills the gap between the ground state excitons and the interband
continuum in the linear absorption spectra\cite{key-2}. As the temperature
drops to 10K, the peak ``A\textquotedblright{} and ``B\textquotedblright{}
are both blue-shifted by around 0.1eV and peak ``C\textquotedblright{}
is shifted by 0.06eV as shown in Figure 1c. The difference of the
blue-shift is the direct consequence of the diverse locations of the
excitons in the Brillouin zone: exciton ``A\textquotedblright{} and
``B\textquotedblright{} are formed at K valley while ``C\textquotedblright{}
is around \foreignlanguage{greek}{$\Gamma$}\inputencoding{latin9}
point. Nevertheless, the continuous absorption still survives and
no distinct single-particle band edge emerges at cryogenic temperature
(10K). The linear absorption spectra cannot resolve the exciton binding
energy. 

\begin{figure}
\includegraphics[width=8.7cm]{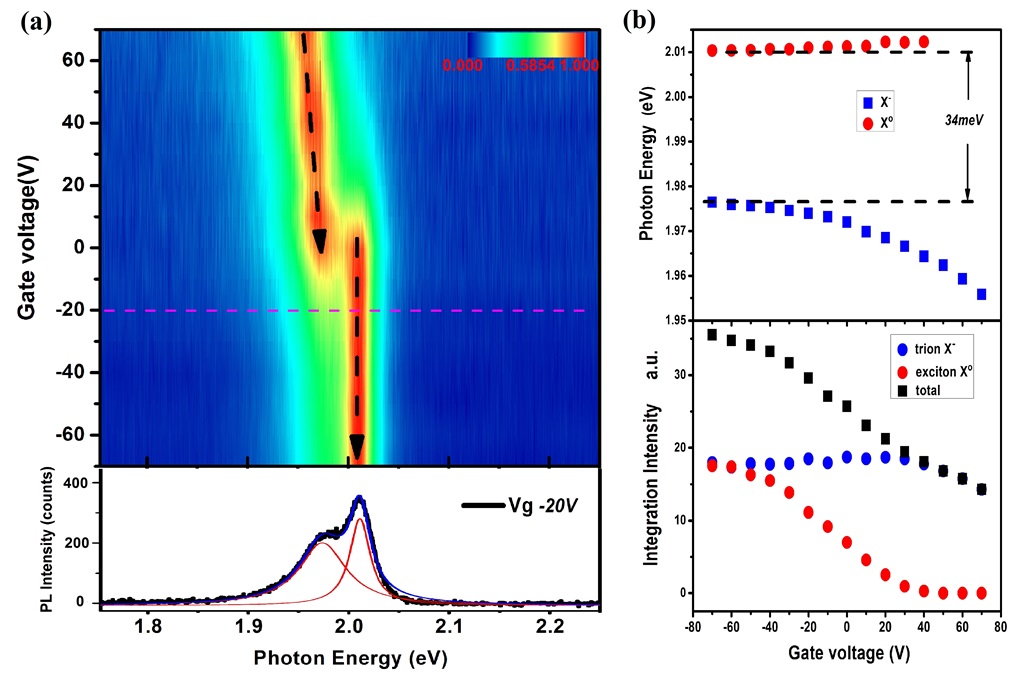}

\caption{Electric doping dependent PL spectra at room temperature. (a) Colour
contour plot of normalized PL spectra excited by a cw laser (2.331eV)
under various back-gate bias. Dashed black arrows contour PL peaks
of free exciton and trion states. Even at zero gate bias the trion
X\textsuperscript{-} exists due to defects or substrate interactions.
Fig. 2(b) illustrates the PL profile at Vg=-20V (dashed line labelled
in the top panel) as a superposition of two Lorentzian shape lines
in red. (c) Electric gating dependence of excitons' and trions'
energies (upper panel) and the corresponding integrated intensities
(bottom panel).}
\end{figure}

Figure 1b shows the absorbance of WS\textsubscript{2} atomically
thin films as a function of the thickness above their ground exciton
energy, which is approximated with the differential transmission.
The absorbance of monolayer and multilayers is linearly proportional
to their thickness, each layer absorbing around 2.0\% and 3.4\% at
excitations of 2.088eV and 2.331eV respectively. The linear layer
dependence of the absorption gives an experimental evidence of the
suppression of interlayer hopping in \textit{2H} stacked WS\textsubscript{2}
as a result of spin-valley coupling\cite{key-5,key-19}. The thickness
dependence could also be used as a thickness monitor for multilayer/monolayer
characterization. There is a side bump at the red side of exciton
A, which modifies the lineshape away from the symmetric Lorentzian
or Gaussian shape. We tentatively attribute the bump to the effect
of electron/hole bound exciton or trion\cite{trion-mose2,trion-WS2,trion_mos2}.
Although the monolayer WS\textsubscript{2} is not intentionally doped,
the structural defects and substrate effects such as charge transfer
and defects modulate the carrier density away from its insulating
state. To confirm the origin of the bump around exciton A, we record
the PL spectra of monolayer WS\textsubscript{2} at various electric
gating (from 70V to -70V) which continuously tunes the Fermi level
of monolayer WS\textsubscript{2}\textsubscript{} as illustrated
in figure 2a.

There is a prominent peak X\textsuperscript{-} at the red side of
the free exciton X\textsuperscript{0} at $V_{g}\approx-20V$ and
the PL spectrum could be described by a superposition of two Lorentzian
curves which center at peak X\textsuperscript{0} and X\textsuperscript{-}
respectively as illustrated in figure 2b. As the gate voltage goes
towards positive values (\textit{Vg>0}), the free exciton X\textsuperscript{0}
gradually diminishes and disappears at \textit{Vg>40V}. Meanwhile
the red-side X\textsuperscript{-} rises to take over the overwhelming
weight of the whole PL until starts to decrease at \textit{Vg>20V}
probably due to the electrostatic screening effect\cite{key-20},
and the peak X\textsuperscript{-} is further red-shifted. The electric
gating dependence attributes X\textsuperscript{-} to \textit{n}-type
trion (electron-bounded exciton) states. As \textit{Vg} goes to negative
bias , the free exciton state X\textsuperscript{0} takes over the
weight of the PL and tends to saturate around \textit{Vg=-70V}. While
the trion state X\textsuperscript{-} monotonically diminishes, the
redshift also shows a sign of saturation of -34meV at around Vg=-70V.
This confirms the trion (electron-bound exciton) origin of the side
bump around exciton A in the monolayer transmission spectrum and the
trion binding energy of 34meV in monolayer. If we follow the simplified
trion model in conventional quantum wells\cite{trion-model} and take
the effective mass of either m\textsubscript{e}=0.37 and m\textsubscript{h}=-0.48\cite{key-21}
or m\textsubscript{e}=0.27,m\textsubscript{h}=-0.32\cite{key-7},
the binding energy of free exciton is estimated at $E_{b}\approx0.1eV$.

\begin{figure}
\includegraphics[width=9cm]{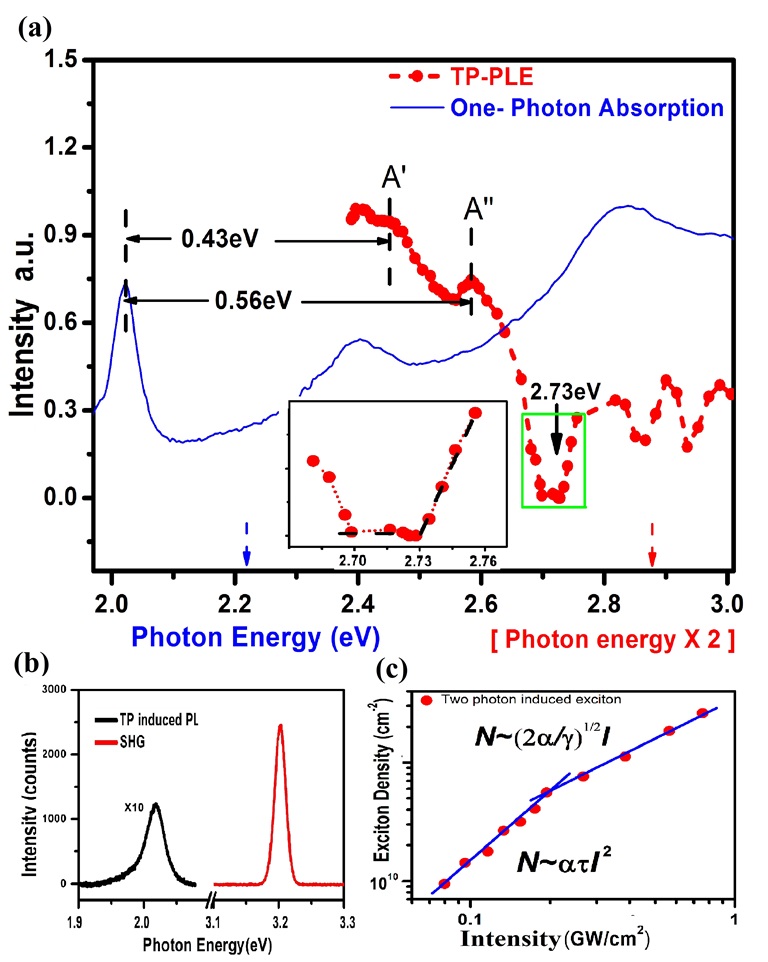}

\caption{(a) One-photon absorption spectrum (blue) in visible range and two-photon
photoluminescence exciton spectrum (red) with the excitation in the
range of 1.192$\thicksim$1.5eV where the \textit{x} axis presents
the exciton energy (in blue) for linear absorption and the double
of the excitation energy (in red) for TP-PLE. A' and A\textquotedblright{}
denote the excited states of exciton A and the zoom-in of the gap
state section is shown in inset. The TP-PL intensity linearly increases
with the excitation energy just above the threshold of the interband
continuum, presenting the signature of two-photon process in 2D systems
where the polarization sits in the 2D plane. (b) The spectra of TP-PL
and the second harmonic generation (SHG) at the excitation of 1.6eV.
The integrated intensity of the PL is more than one order of magnitude
less than that of the SHG. (c) The intensity of ground state exciton
\textit{vs.} the excitation intensity under the excitation of \textit{1.59eV}.
The fitting lines demonstrate a quadratic- (under low intensity) and
a linear-dependence (under high intensity) respectively, which yield
the exciton-exciton annihilation rate $\gamma\approx0.31-0.47cm^{2}/s$
and the two-photon absorption cross section $\alpha\thickapprox3.5-5.3\times10^{4}cm^{2}W^{-2}S^{-1}$. }
\end{figure}

Two-photon excitation is a third order optical process involving simultaneous
absorption of the two photons which follows selection rules different
from those in one-photon (linear) process. As a photon has an odd
intrinsic parity, one- and two-photon transitions are mutually exclusive
in systems with inversion symmetry : one-photon transitions are allowed
between states with different parity and two-photon transitions between
states with the same parity; in systems without inversion symmetry
like monolayer TMDCs described by a point group of D\textsubscript{3h}
symmetry, parity is not a good quantum number and there exist transitions
which are both one- and two-photon allowed. Nevertheless, the oscillator
strengths of exciton states are generally different between one- and
two-photon processes. A simplified exciton model could be described
as $U_{n}^{l}(\rho=r_{e}-r_{h})\phi_{c}(r_{e})\phi_{v}(r_{h})$ where
$\phi_{c}(r_{e})$($\phi_{v}(r_{h}))$ presents the electron (hole)
wave function, and $U_{n}^{l}$ is the function of relative motion
of electron-hole. The optical transition rates for one- and two-photon
processes\cite{two-photon} 
\[
\begin{array}{cc}
W_{OP}\sim\mid A\mid^{2}\underset{c,v}{\sum}\left|\left\langle c\left|\varepsilon\cdot p\right|v\right\rangle \right|^{2}\underset{c,v}{\sum}\left|\left\langle \phi_{c}\left|\phi_{v}\right.\right\rangle \right|^{2}\\
\cdot\left|U_{n}^{l}(\rho=0)\right|^{2}S_{cv}(\hbar\omega)
\end{array}
\]

\begin{eqnarray*}
 & \begin{array}{cc}
W_{TP}\sim\mid A_{1}A_{2}\mid^{2}\underset{c,v}{\sum}\left|\left\langle c\left|\varepsilon\cdot p\right|v\right\rangle \right|^{2}\underset{c,v}{\sum}\left|\left\langle \phi_{c}\left|\phi_{v}\right.\right\rangle \right|^{2}\\
\left|\nabla U_{n}^{l}(\rho=0)\right|^{2}S_{cv}(\hbar\omega_{1}+\hbar\omega_{2})
\end{array}
\end{eqnarray*}
where \textit{A} denotes the vector potential of the excitation, $\varepsilon$
the light polarization unit vector, $<c\left|\varepsilon\cdot p\right|v>$
the interband matrix elements, and $S_{cv}(\hbar\omega)$ the line-shape
function of interband exciton. In a 2D system,$U_{n}^{l}$ could be
described by a solution of 2D Wannier-Mott exciton $U_{n}^{l}(\rho,\theta)=\frac{1}{\sqrt{\pi}(n-\frac{1}{2})^{3/2}}\sqrt{\frac{(n-l)!}{(n+l)!}}\left(\frac{2\rho}{n-\frac{1}{2}}\right)^{l}exp\left(-\frac{\rho}{n-\frac{1}{2}}\right)L_{n}^{2l}(\frac{2\rho}{n-\frac{1}{2}})exp\left(il\theta\right)$
and the exciton binding energy could be described as $E_{n}=\frac{Ry^{*}}{(n-\frac{1}{2})^{2}}$
where\textit{ n}=1,2\ldots{} is the principle quantum number, \textit{l}=0,1,..(\textit{n}-1)
is the angular quantum number, and $L_{2n}^{l}$ is the associate
Laguerre polynomial. As the exciton oscillator strength decays as
$n^{-3}$, only the ground state (\textit{n}=1) and the first two
excited states are considered. In a one-photon process, the ground
state \textit{1s} (\textit{n}=1,\textit{l}=0) dominates; Whereas in
a two-photo process the ground state and \textit{ns} states (\textit{l}=0)
are dramatically subsidized owing to $\nabla U_{n}^{l}(\rho=0)\approx0$
and the \textit{p} state dominates. Analyzing the difference between
one- and two-photon processes would lead to extracting the exciton
binding energy of monolayer TMDC.

Figure 3 shows a TP-PLE spectrum of monolayer WS\textsubscript{2},
where the PL intensity of free band-edge exciton A is recorded as
a function of the pulsed excitation energy. With the contrasting optical
transition strength, two-photon excitation resonant with p type exciton
states dominates while s type is nearly invisible. The prominent PL
occurs at the excitation around 1.2eV which is the half of the exciton
B energy and 1.29eV. There is a significant gap state in the range
of 1.35-1.36eV where the PL intensity drops to nearly negligible.
The negligible but nonzero PL intensity likely results from the re-absorption
of the second harmonic excitation, since the SHG intensity is more
than one order of magnitude higher than that of two-photon luminescence
as shown in Figure 3b. Upon the excitation just above the gap (> 1.365eV)
as indicated by the arrow in Figure 3a, the PL intensity shows a linear
increase with the excitation energy as indicated in the inset. It
is the signature of two-photon absorption with in-plane polarization
in 2D system\cite{two-photon}. Besides, the two local minimums at
higher excitation energy around 1.44eV and 1.46eV have significant
PL intensity and therefore are unlikely to be the single-particle
band gap state. Thus the single-particle gap could be determined at
2.73eV (2X1.365eV), consistent with Ref\cite{Ye}. Given that the
PL peaks at 2.02eV presenting the energy of the\textit{ }ground-state
exciton, the exciton binding energy of $E_{b}=0.71\pm0.01eV$ is extracted
from the energy difference between the ground-state exciton and the
onset of the inter-band continuum.

With the band-edge exciton binding energy of 0.71eV we could attribute
the peaks around 2.42eV (2X1.21eV) and 2.58eV (2X1.29eV) in the TP-PLE
spectrum to the excited states of excitons, which are qualitatively
consistent with the recent \textit{ab initio} calculation\cite{Ye}.
As the exciton A and B both originate from the spin-split valence
bands at K(K') valley with the similar effective mass, a similar
strength of binding energy is expected. Besides, the PL intensity
around the peak A' and A'' monotonically decreases. Both peaks are
likely to be the excited state of the same exciton, and we tentatively
attribute peaks A' and A\textquotedblright{} to the \textit{2p} and
\textit{3p} states of exciton A respectively. The exciton binding
energy could also be evaluated from the energy difference between
exciton \textit{1s }and\textit{ np} states. The 2D hydrogen model
gives the energy difference between \textit{1s} and \textit{2p}(\textit{3p})
\[
E_{b}=4Ry^{*}=\frac{9}{8}\triangle E_{1s-2p}=\frac{100}{96}\triangle E_{1s-3p}
\]
which correspond to $E_{b}=0.48eV$ and $E_{b}=0.58eV$ respectively.
The alternative assignment for example peak A' to 3p state leads to
$E_{b}=0.45eV$. These are significantly smaller than $E_{b}=0.71\pm0.01eV$
extracted from the energy difference between the ground state exciton
and the onset of the inter-band continuum, and the distribution of
these excited states also significantly deviates from that of the
2D hydrogen model. The difference may lie in the modification of the
2D hydrogen model by electron-phonon and electron correlation interactions
in monolayer TMDC. Recent first principle simulation shows that q-dependent
screening dramatically enhances the binding energy of the excited
states of excitons\cite{key-2,Ye}. Nevertheless, it is safe to extract
the exciton binding energy of $E_{b}=0.71\pm0.01eV$ by the energy
difference between 1s exciton and the onset of the interband continuum,
independent of the assignments of the excited states. It also implies
that the model to estimate the exciton binding energy via trion binding
energy is inappropriate.

The two-photon absorption has a quadratic dependence on the excitation
intensity in principle. The two-photon photoluminescence (TP-PL) intensity
from monolayer WS\textsubscript{2} displays a clear quadratic dependence
on the excitation intensity at low power as shown in figure 3c. As
the excitation intensity increases above $0.2GW/cm^{2}$, the PL intensity
experiences a clear transition from quadratic to linear dependence
on the excitation intensity. If we follow the simple model 
\[
\frac{dN}{dt}=\alpha I^{2}-\frac{N}{\tau}-\frac{1}{2}\gamma N^{2}=0
\]
where \textit{N} denotes the exciton density, \textit{I }the excitation
intensity, $\alpha$ the two-photon absorption cross section, $\tau$
the exciton lifetime and $\gamma$ the exciton-exciton annihilation
rate, the fitting of the quadratic dependence $I_{ph}=\alpha\tau I^{2}$
($I\rightarrow0$) gives two-photon absorption cross section of $\alpha\thickapprox3.5-5.3\times10^{4}cm^{2}W^{-2}S^{-1}$
at 1.59eV where the PL quantum yield of $4\times10^{-3}$\cite{key-3}
and the exciton lifetime of \textit{100ps} are assumed\cite{JF}.
Subsequently the linear dependence slope at high intensity $\sqrt{\frac{2\alpha}{\gamma}}$
yields the exciton-exciton annihilation rate $\gamma\approx0.31-0.47cm^{2}/s$
which is qualitatively consistent with that in monolayer MoSe$_{2}$
measured by pump-probe reflection spectroscopy\cite{zhaohui}. The
linear intensity dependence of TP-PL is the evidence of the strong
exciton-exciton interactions in monolayer TMDCs.

In summary, the linear absorption spectroscopy cannot resolve the
electronic interband transition edge down to 10K due to the strong
electron-phonon scattering and the overlap of excitons around $\Gamma$
point. The TP-PLE measurements successfully probe the excited states
of the band-edge exciton and the single-particle band gap. The exciton
binding energy of 0.71$\pm$0.01eV is extracted by the energy difference
between 1s exciton and the single-particle gap in monolayer WS$_{2}$.
The distribution of the exciton excited states significantly deviates
from the 2D hydrogen model. The giant exciton binding energy manifests
the unprecedented strong Coulomb interactions in monolayer TMDCs. 

\begin{acknowledgments}
The work is supported by Area of excellency (AoE/P-04/08 ), CRF of
Hong Kong Research Grant Council and SRT on New Materials of University
of Hong Kong \end{acknowledgments}

\end{document}